\documentstyle[amssymb,preprint,aps]{revtex}

\begin{document}
\title{{\large {\bf Supersymmetric Inflation, Baryogenesis and $\nu_{\mu} -
\nu_{\tau}$ Oscillations}}}
\author{{\bf G. Lazarides $^{a}$}, {\bf Q. Shafi $^{b}$} and {\bf N. D. Vlachos $%
^{c} $}}
\address{a) Theory Division, CERN, 1211 Geneva 23,\\
Switzerland \\
and \\
Physics Division, School of Technology, Aristotle University of Thessaloniki,%
\\
Thessaloniki GR 540 06, Greece.\\
b) Bartol Research Institute, University of Delaware,\\
Newark, DE 19716, USA. \\
c) Department of Physics, Aristotle University of Thessaloniki,\\
Thessaloniki GR 540 06, Greece.}
\maketitle

\begin{abstract}
In a supersymmetric left-right symmetric model, inflation, baryogenesis (via
leptogenesis) and neutrino oscillations can become closely linked. A
familiar ansatz for the neutrino Dirac masses and mixing of the two heaviest
families, together with the MSW resolution of the solar neutrino puzzle,
imply that $1\ {\rm {eV}\stackrel{_{<}}{_{\sim }}m_{\nu _{\tau }}\stackrel{%
_{<}}{_{\sim }}9\ eV}$. The predicted range for the mixing angle $\theta
_{\mu \tau }$ will be partially tested by the Chorus/Nomad experiment. The
CP violating phase $\delta _{\mu \tau }$ is also discussed.
\end{abstract}

\nopagebreak
\vskip0.5truecm

The minimal supersymmetric standard model (MSSM) provides a particularly
compelling extension of the $SU(3)_{c}\times SU(2)_{L}\times U(1)_{Y}$ gauge
theory. Yet, it seems quite clear that MSSM, in turn, must be part of a
larger picture. Let us list some reasons why: i) In MSSM, there is no
understanding of how the supersymmetric $\mu $ term is $\sim 10^{2}$-$10^{3}$
GeV. In principle, it could be as large as the Planck mass. ii) An important
(and undetermined) new parameter in MSSM is $\tan \beta $, the ratio of the
vacuum expectation values of the two higgs doublets. Among other things, an
understanding of this parameter can shed light on the mass of the
Weinberg-Salam higgs. iii) It has become increasingly clear that a
combination of both `cold' and `hot' dark matter (CHDM) provides [1] a good
fit to the data on large scale structure formation, especially if the
primordial density fluctuations are essentially scale invariant. In MSSM,
there is no HDM candidate, even after including non-renormalizable terms.
iv) It has been impossible, so far, to implement inflation within MSSM. v)
Last, but not least, it is not easy to generate in MSSM the observed baryon
asymmetry through the electroweak sphaleron processes.

Remarkably, all these challenges can be overcome in one fell swoop by
considering a modest extension of the MSSM gauge symmetry to $H\equiv
SU(3)_{c}\times SU(2)_{L}\times SU(2)_{R}\times U(1)_{B-L}$. Of course, it
is anticipated that $H$ is embedded in a grand unified theory such as $%
SO(10) $ or $SU(3)_{c}\times SU(3)_{L}\times SU(3)_{R}$. Apart from
aesthetics, there are tantalizing hints from the extrapolation of low energy
data for the existence of a supersymmetric unification scale $\sim 10^{16}$
GeV. The details on how the extension of MSSM to an $H$-based model can
resolve the points above will not be discussed here, especially since an
inflationary scenario based on $H$ has been considered in some detail
elsewhere [2,3]. This scenario gives rise to an essentially scale invariant
spectrum (spectral index $n\simeq 0.98$), and contains both `cold' (LSP) and
`hot' (massive neutrinos) dark matter candidates. The observed baryon
asymmetry is generated through partial conversion of a primordial lepton
asymmetry [4]. Finally, the parameter $\tan \beta $ is close to $m_{t}/m_{b}$%
, which also explains why the higgs boson of the electroweak theory has not
been seen at LEP II. Its tree level mass is $M_{Z}$ which, after radiative
corrections, becomes $m_{h^{\circ }}\simeq 105-120$ GeV.

The inflationary phase is associated with the gauge symmetry breaking $%
SU(2)_{R}\times U(1)_{B-L}\rightarrow U(1)_{Y}$. Of course, since $H$ is
presumably embedded in some grand unified symmetry, there may well be more
than one inflationary epoch. We concentrate on the last and most relevant
one. The above breaking is achieved by a pair of $SU(2)_{R}$ doublet `higgs'
superfields which have the same gauge quantum numbers as the `matter' right
handed neutrino superfields. As a consequence, the inflaton decays primarily
into `matter' right handed neutrinos via quartic (or higher order)
superpotential couplings. The `reheat' temperature, $T_{R}$, turns out to be
about one order of magnitude smaller than the mass of the heaviest right
handed neutrino that the inflaton can decay into. The gravitino constraint
on $T_{R}\,$($\lesssim 10^{9}$ GeV) allows us to restrict the second and
third family right handed neutrino masses $M_{2},M_{3}$ in a fairly narrow
range. Our approach poses no obvious constraint on the first family right
handed neutrino mass $M_{1}$, except from $M_{1}\leq M_{2}$. The constraints
on $M_{2},M_{3}$, however, together with the leptogenesis scenario will
enable us to restrict the oscillation parameters of the $\nu _{\mu }$-$\nu
_{\tau }$ system [5].

We consider the $2\times 2$ `asymptotic' mass matrices $M^{\text{L}}$, $M^{%
\text{D}}$ and $M^{\text{R}}$ in the weak basis, where the superscripts L,D
and R denote the charged lepton, neutral Dirac, and right handed neutrino
sectors respectively. $M^{\text{L}}$,$M^{\text{D}}$ are diagonalized by the
biunitary rotations $L=U^{L}L^{\prime }$,$\ L^{c}=U^{L^{c}}L^{c}\,^{\prime }$%
,$\,\nu =U^{\nu }\nu ^{\prime }$, $\nu ^{c}=U^{\nu ^{c}}\nu ^{c}\,^{\prime }$%
: 
\begin{equation}
M^{L}\rightarrow M^{L}\,^{\prime }=\tilde{U}^{L^{c}}M^{L}U^{L}=\left( 
\begin{array}{cc}
m_{\mu } &  \\ 
& m_{\tau }
\end{array}
\right) \ \ ,
\end{equation}
\begin{equation}
M^{D}\rightarrow M^{D}\,^{\prime }=\tilde{U}^{\nu ^{c}}M^{D}U^{\nu }=\left( 
\begin{array}{cc}
m_{2}^{D} &  \\ 
& m_{3}^{D}
\end{array}
\right) \ ,
\end{equation}
where the diagonal entries are positive. This gives rise to the `Dirac'
mixing matrix $U^{\nu }\,^{\dagger }U^{L}$ in the leptonic charged currents.
Using the remaining freedom to perform arbitrary phase transformations on
the components of $L^{\prime }$, $\nu ^{\prime }$ together with the
compensating ones on the components of $L^{c}\,^{\prime }$,$\nu
^{c}\,^{\prime }$ so that $M^{L}\,^{\prime }$,$M^{D}\,^{\prime }$ remain
unaltered, we can bring this matrix to the form 
\begin{equation}
U^{\nu }\,^{\dagger }U^{L}\rightarrow \left( 
\begin{array}{cc}
{\cos \theta }^{D} & {\sin \theta }^{D} \\ 
-{\sin \theta }^{D} & {\cos \theta }^{D}
\end{array}
\right) ,
\end{equation}
where $\theta ^{D}(0\leq \theta ^{D}\leq \pi /2)$ is the `Dirac' (not the
physical) mixing angle in the $2$-$3$ leptonic sector.

In this basis, depicted with a double prime on the superfields, the Majorana
mass matrix can be written as: 
\begin{equation}
M^{R}=U^{-1}\ M_{0}\ \tilde{U}^{-1}\ ,
\end{equation}
where $M_{0}=diag(M_{2},M_{3})$, with $M_{2},M_{3}$ (both positive) being
the two Majorana masses, and $U$ is a unitary matrix which can be
parametrized as 
\begin{equation}
U=\left( 
\begin{array}{cc}
{\cos }\theta & {\sin }\theta \ e^{-i\delta } \\ 
-{\sin }\theta \,e^{i\delta } & {\cos }\theta
\end{array}
\right) \left( 
\begin{array}{cc}
e^{i\alpha _{2}} &  \\ 
& e^{i\alpha _{3}}
\end{array}
\right) \ ,
\end{equation}
with $0\leq \theta \leq \pi /2$ and $0\leq \delta <\pi $. The light neutrino
mass matrix, to leading order in $M^{R}\,^{-1}M^{D}\,^{\prime }$, is 
\begin{equation}
m=-\tilde{M}^{D}\,^{\prime }\ \frac{1}{M^{R}}\ M^{D}\,^{\prime }\ \ ,
\end{equation}
where $M^{D}\,^{\prime }$ is defined in eq.(2). We can express $m$ as 
\begin{equation}
m=\left( 
\begin{array}{cc}
e^{i\alpha _{2}} &  \\ 
& e^{i\alpha _{3}}
\end{array}
\right) \Psi (\theta ,\delta )\left( 
\begin{array}{cc}
e^{i\alpha _{2}} &  \\ 
& e^{i\alpha _{3}}
\end{array}
\right) \ \ ,
\end{equation}
where $\Psi (\theta ,\delta )$ depends also on $M_{2}$, $M_{3}$, $m_{2}^{D}$%
, $m_{3}^{D}$ $.$ We diagonalize $m$ by a unitary rotation $\nu ^{\prime
\prime }=V\nu ^{\prime \prime \prime }$ with 
\begin{equation}
V\ =\ \left( 
\begin{array}{cc}
e^{i\beta _{2}} &  \\ 
& e^{i\beta _{3}}
\end{array}
\right) \left( 
\begin{array}{cc}
{\cos }\varphi & {\sin }\varphi \,e^{-i\epsilon } \\ 
-{\sin }\varphi \,e^{i\epsilon } & {\cos }\varphi
\end{array}
\right) \ ,\ 
\end{equation}
where $0\leq \varphi \leq \pi /2\ ,\ 0\leq \epsilon <\pi $.The `Dirac'
mixing matrix in eq.(3) is now multiplied by $V^{\dagger }$ on the left and,
after suitable phase absorptions, takes the form 
\begin{equation}
\left( 
\begin{array}{cc}
{\cos }\theta _{23} & {\sin }\theta _{23}\,e^{-i\delta _{23}} \\ 
-{\sin }\theta _{23}\,e^{i\delta _{23}} & {\cos }\theta _{23}
\end{array}
\right) \ ,\ 
\end{equation}
where $0\leq \theta _{23}\leq \pi /2\,\ ,\ 0\leq \delta _{23}<\pi $. Here, $%
\theta _{23}$ (or $\theta _{\mu \tau }$) is the physical mixing angle in the 
$2$-$3$ leptonic sector and its cosine equals the modulus of the complex
number 
\begin{equation}
{\cos }\varphi \,{\rm {\cos }\theta }^{D}+{\rm {\sin }\varphi \sin \theta }%
^{D}\,e^{i(\xi -\epsilon )}\ ,{\rm \ }
\end{equation}
where $-\pi \leq \xi -\epsilon =\beta _{2}-\beta _{3}-\epsilon \leq \pi $.
Moreover, $\delta _{23}$ (or $\delta _{\mu \tau }$) is the associated CP
violating phase which is given by $\delta _{23}=\xi +\rho -\lambda $ (modulo 
$\pi $), where $\lambda \ (-\pi \leq \lambda \leq \pi )$ and $-\rho \ (-\pi
\leq \rho \leq \pi )$ are the arguments of the complex numbers in eq.(10)
and ${\cos }\varphi \,{\rm {\sin }}\theta ^{D}-{\rm {\sin }}\varphi \,{\rm {%
\cos }}\theta ^{D}\,e^{i(\xi -\epsilon )}$ respectively. Since $\xi $
remains undetermined (see below), the precise values of $\theta _{23}$ and $%
\delta _{23}$ cannot be found. However,we can determine the range in which $%
\theta _{23}$ lies: 
\begin{equation}
|\,\varphi -\theta ^{D}|\leq \theta _{23}\leq \varphi +\theta ^{D},\ {\rm {%
for}\ \varphi +\theta }^{D}\leq \ \pi /2\,\cdot
\end{equation}
The double valued function $\delta _{23}(\theta _{23})$ is also determined.

We will denote the two positive eigenvalues of the light neutrino mass
matrix by $m_{2}$ (or $m_{\nu _{\mu }}$) , $m_{3}$ (or $m_{\nu _{\tau }}$)
with $m_{2}\leq m_{3}$. Recall that all the quantities here (masses,
mixings, etc.) are `asymptotic' (defined at the GUT scale). The determinant
and the trace invariance of $\Psi (\theta ,\delta )^{\dagger }\Psi (\theta
,\delta )$ provide us with two constraints on the (asymptotic) parameters
which take the form: 
\begin{equation}
m_{2}m_{3}\ =\ \frac{\left( m_{2}^{D}m_{3}^{D}\right) ^{2}}{M_{2}\ M_{3}}\ \
,
\end{equation}
\begin{equation}
m_{2}\,^{2}+m_{3}\,^{2}\ =\frac{\left( m_{2}^{D}\,\,^{2}{\rm c}%
^{2}+m_{3}^{D}\,^{2}{\rm s}^{2}\right) ^{2}}{M_{2}\,^{2}}+
\end{equation}
\[
\ \frac{\left( m_{3}^{D}\,^{2}{\rm c}^{2}+m_{2}^{D}\,^{2}{\rm s}^{2}\right)
^{2}}{M_{3}\,^{2}}+\ \frac{2(m_{3}^{D}\,^{2}-m_{2}^{D}\,^{2})^{2}{\rm c}^{2}%
{\rm s}^{2}\,{\cos 2\delta }}{M_{2}\,M_{3}}, 
\]
where $\theta $, $\delta $ are defined in eq.(5) , ${\rm c}=\cos \theta \,,\ 
{\rm s}=\sin \theta $.

We now need information about the quantities $m_{2}^{D}$,$m_{3}^{D}$ (the
`asymptotic' Dirac masses of the muon and tau neutrinos) as well as the
`Dirac' mixing angle $\theta ^{D}$. A plausible assumption, inspired by $%
SO(10)$ for instance, is that these quantities are related to the quark
sector parameters by the $SU(4)_{c}$ symmetry. In other words, we will
assume the asymptotic relations: 
\begin{equation}
m_{2}^{D}=m_{c}\ ,\ m_{3}^{D}=\ m_{t}\ ,\ {\sin }\theta ^{D}=|V_{cb}|\quad
\cdot
\end{equation}
Of course, the $SU(4)_{c}$ symmetry is not expected to hold in the down
sector of the second family.

Contact with experiment can be made after renormalization effects have been
taken into account. The pair of MSSM higgs doublets is assumed to belong to
the $(2,2)$ representation of $SU(2)_{L}\times SU(2)_{R}$, implying that $%
\tan \beta \simeq m_{t}/m_{b}$ both `asymptotically' and at low energies.
The light neutrino masses, in this case, can be obtained by dividing the
right hand side of eq.(6) by a factor of 2.44 [6]. Eqs.(12),(13) now hold
with $m_{2}$,$m_{3}$ being the low energy neutrino masses and $m^{D}$ 's
replaced by their asymptotic values divided by 1.56. The latter turn out to
be $m_{2}^{D}\simeq \ 0.23$ GeV and $\ m_{3}^{D}\simeq \ 116$ GeV (with ${%
\tan }\beta \ \simeq \ m_{t}/m_{b}$)[7]. Finally, using $SU(4)_{c}$
invariance, the asymptotic `Dirac' mixing angle is calculated from ${\sin }%
\theta ^{D}=\ |V_{cb}|{\rm \ }$(asymptotic) $\simeq 0.03$. The
renormalization of the mixing angle $\theta _{23}$, with ${\tan }\beta \
\simeq \ m_{t}/m_{b}$, has been considered in ref.[6]. The net effect is
that ${\sin }^{2}2\theta _{23}$ increases by about 40\% from $M_{GUT}$ to $%
M_{Z}$.

In view of the lack of a compelling alternative theoretical framework, we
will assume the hierarchy $m_{1}\ll m_{2}\ll m_{3}$. We will thus restrict $%
m_{2}$ in the range $1.7\times 10^{-3}\ $eV${\rm \ }\lesssim \ m_{2}\
\lesssim \ 3.5\times 10^{-3}$ eV, as allowed by the small angle MSW solution
[8]. We now recall a few salient features of the inflationary scenario
associated with the breaking $SU(2)_{R}\times U(1)_{B-L}\rightarrow U(1)_{Y}$%
. As the inflaton ($SU(2)_{R}$ doublets $\phi ,\bar{\phi}$) oscillates about
its minimum, it decays into the appropriate `matter' right handed neutrino ($%
\nu ^{c}$) via the effective superpotential coupling $\nu ^{c}\nu ^{c}\bar{%
\phi}\bar{\phi}$ permitted by the gauge symmetry. The `reheat' temperature $%
T_{R}$ is then related [3] to the mass $M_{H}$ of the heaviest right handed
neutrino the inflaton can decay into: $T_{R}\ \simeq \ M_{H}/9.2$.The
inflaton mass is given by $m_{infl}\ \simeq \ 3.4\times 10^{13}$ GeV. If $%
M_{2}$,$M_{3}$ are smaller than $m_{infl}/2$, the inflaton decays
predominantly into the heaviest of the two. Then, eq.(12) and the
cosmological bound $m_{3}\lesssim 23$ eV [9] require the smallest allowed
mass of the heaviest right handed neutrino to be $\simeq 9.4\times 10^{10}$
GeV giving $T_{R}\gtrsim \ 10^{10}$ GeV, in clear conflict with the
gravitino constraint. Consequently, we find that 
\begin{equation}
m_{infl}/2\ \leq \ M_{3}\lesssim \ 2.5\times 10^{13}\ {\rm GeV,}
\end{equation}
where the upper bound comes from the requirement that the coupling constant
of the non-renormalizable superpotential term $\nu ^{c}\nu ^{c}\bar{\phi}%
\bar{\phi}$, which provides mass for the right handed neutrinos, should not
exceed unity. Thus, $M_{3}$ is restricted in a narrow range, and the
inflaton decays into the second heaviest right handed neutrino.

The lepton asymmetry is generated by the subsequent decay of this neutrino
and is given by [4] 
\begin{equation}
\frac{n_{L}}{s}\ =-\ \frac{9\,T_{R}}{8\pi \,m_{infl}}\ \frac{M_{2}}{M_{3}}\ 
\frac{Im(UM^{D}\,^{\prime }M^{D}\,^{\prime }\,^{\dagger }U^{\dagger
})_{23}^{2}}{v^{2}(UM^{D}\,^{\prime }M^{D}\,^{\prime }\,^{\dagger
}U^{\dagger })_{22}}\ \ ,
\end{equation}
where $v$ is the electroweak VEV at $M_{GUT}$. Substituting $U$ from eq.(5),
we get 
\begin{equation}
\frac{n_{L}}{s}=\frac{9\,T_{R}}{8\pi \,m_{infl}}\,\frac{M_{2}}{M_{3}}\,\frac{%
{\rm c}^{2}{\rm s}^{2}\ \sin 2\delta \ (m_{3}^{D}\,^{2}-m_{2}^{D}\,^{2})^{2}%
}{v^{2}(m_{3}^{D}\,^{2}\ {\rm s}^{2}\ +\ m_{2}^{D}\,^{2}{\rm \ c^{2}})}\
\cdot
\end{equation}
Here we can again replace $m_{2}^{D}$, $m_{3}^{D}\,$ by their `asymptotic'
values divided by 1.56 and $v$ by $174$ GeV. Assuming the MSSM spectrum
between $1$ TeV and $M_{GUT}$, the observed baryon asymmetry $n_{B}/s$ is
related [10] to $n_{L}/s$ by $n_{B}/s\ =-28/79\,(n_{L}/s).$

We take a fixed value of $M_{3}$ in the allowed range (15) and, for any pair
of values $(m_{2},\ m_{3})$, we calculate $M_{2}$ and $T_{R}$. The
constraint $T_{R}\ \leq \ 10^{9}$ GeV yields a lower bound for the product $%
m_{2}m_{3}$ excluding the region below a hyperbola on the $m_{2}$,$m_{3}$
plot. Note that the gravitino constraint combined with the MSW restriction
on $m_{2}$ yields a lower bound for $m_{3}$. Inside the allowed $m_{2}$,$%
m_{3}$ region, we can use the trace condition (13) to solve for $\delta
(\theta )$ in the interval $0\ \leq \ \theta \ \leq \ \pi /2$. Given that we
need a negative value for $n_{L}/s$ so that $n_{B}/s>0$, we see that there
is at most one useful branch of the function $\delta (\theta )$ taken to lie
in the region $-\pi /2\leq \delta (\theta )\ \leq 0$ . The expression for $%
\delta (\theta )$ is subsequently substituted in eq.(17) for the leptonic
asymmetry, and the range of $\theta $ satisfying the constraint $%
0.02\lesssim \ \Omega _{B}h^{2}\ \lesssim \ 0.03$ is found. (This constraint
is consistent with the low deuterium abundance as well as with structure
formation in `cold' plus `hot' dark matter models). The $m_{2}$,$m_{3}$
pairs, for which this range of $\theta $ exists, are consistent with the
observed baryon asymmetry. In Fig.1, we depict the areas on the $m_{2}$,$%
m_{3}$ plane which satisfy both the gravitino and baryogenesis constraints
in the two extreme cases of $M_{3}=m_{infl}/2$ (bounded by the thick solid
line) and $M_{3}=2.5\times 10^{13}$ GeV (bounded by the thick dashed line).
The lower line, in both cases, corresponds to the gravitino constraint
whereas the upper one comes from the baryogenesis constraint. Note that the
baryogenesis constraint at $m_{2}=1.7\times 10^{-3}$ eV provides us with an
upper bound for $m_{3}$. In summary, we get an interesting upper as well as
a lower bound on $m_{3}$ (i.e. on $\Omega _{HDM}$!), and, for each $m_{3}$
value, we know the allowed range of $m_{2}$. Namely, for $M_{3}=m_{infl}/2,\
1.3{\ }${eV }$\lesssim m_{3}\lesssim 8.8$ eV whereas, for $M_{3}=2.5\times
10^{13}\ $GeV$,\ 0.9{\ }${eV }$\lesssim m_{3}\lesssim 5.1$ eV. Thus, $m_{\nu
_{\tau }}$ is restricted in the range of 1 to 9 eV.

The discussion above can be extended to yield useful information for $\theta
_{\mu \tau }$. For each allowed pair $m_{2}$,$m_{3}$ and, for every value of 
$\theta $ in the allowed range, we construct $\varphi $ and $\epsilon $ in
eq.(8). The phases $\alpha _{2}$ and $\alpha _{3}$ in eq.(5) remain
undetermined by the conditions (12),(13) and, consequently, $\beta _{2}$,$%
\beta _{3}$ in eq.(8) and $\xi $ in eq.(10) remain also undetermined. This
fact does not allow us to predict the value of $\theta _{\mu \tau }$ for
each value of $\theta $, but only its allowed range given in eq.(11). The
union of all these intervals, comprising all allowed values of $\theta $ and 
$m_{2}$ for a given $m_{3}$, constitutes the allowed range of $\theta _{\mu
\tau }$ for this $m_{3}$. These ranges, after taking the renormalization of $%
\theta _{\mu \tau }$ into account, are depicted in Fig.2 for all possible
values of $m_{3}$, and constitute the allowed area of the neutrino
oscillation parameters. The region bounded by the thick solid line is the
allowed area for $M_{3}=m_{infl}/2$, whereas the one bounded by the thick
dashed line corresponds to $M_{3}=2.5\times 10^{13}$ GeV. The areas tested
(to be tested) by past (future) experiments are also indicated in Fig.2. The
area excluded by E531 is depicted in Fig.1 and lies above the thin solid
(dashed) E531 line, for $M_{3}=m_{infl}/2$ ($2.5\times 10^{13}$ GeV). CDHS
does not appear to have any appreciable effect. Furthermore, if CHORUS gives
negative results, we must further exclude the area above the corresponding
thin solid (dashed) line in Fig.1, for $M_{3}=m_{infl}/2$ ($2.5\times
10^{13} $ GeV). A possibly negative CHORUS result implies that the upper
bound for $m_{\nu _{\tau }}$ drops down to $\simeq 3.7$ eV. Notice that the
upper limit on baryon asymmetry has no effect on Figs.1 and 2.

The CP violating phase $\delta _{\mu \tau }$ as a function of $\theta _{\mu
\tau }$, for given values of $M_{3}$,$m_{2}$,$m_{3}$ and $\Omega _{B}h^{2}$,
can now be constructed. We choose $M_{3}=m_{infl}/2$, $m_{2}=2.6\times
10^{-3}$ eV, $m_{3}=4$ eV, $\Omega _{B}h^{2}=0.025$ and solve for $\theta $.
We find two solutions and, for each one of them and any $\xi \ (-\pi \leq
\xi -\epsilon \leq \pi )$, we calculate $\theta _{\mu \tau }$ and $\delta
_{\mu \tau }=\xi +\rho -\lambda $. Eliminating $\xi $, we obtain the
function $\delta _{\mu \tau }(\theta _{\mu \tau })$ in the region of
eq.(11). This function turns out to be double valued-the sum of the two
branches equals $2\epsilon $- and is depicted in Fig.3, for both values of $%
\theta $, after renormalizing $\theta _{\mu \tau }$. The $\theta _{\mu \tau
} $'s excluded by E531, for $m_{\nu _{\tau }}=4$ eV, are also indicated and
lie to the right of the E531 line.

In conclusion, we find that a modest extension of MSSM to $SU(3)_{c}\times
SU(2)_{L}\times SU(2)_{R}\times U(1)_{B-L}$ can yield significant new
results by tying together a number of apparently unrelated phenomena. In
particular, inflation can be realized, the spectral index $n\simeq 0.98$, we
get both `cold' (essentially bino) and `hot' ($\nu _{\tau }$) dark matter,
while the $\nu _{\tau }$ mass and $\nu _{\mu }$-$\nu _{\tau }$ mixing is
within reach of present and planned experiments. In the simplest scheme, the
atmospheric neutrino anomaly remains a mystery.

We thank K.S. Babu for several discussions regarding the renormalization
effects, and E. Tsesmelis for information about the experiments. TMR and
NATO support under grant numbers ERBFMRXCT-960090 and CRG 970149 is
gratefully acknowledged.

\newpage

\begin{center}
{\large {\bf {Figure Captions}}}
\end{center}

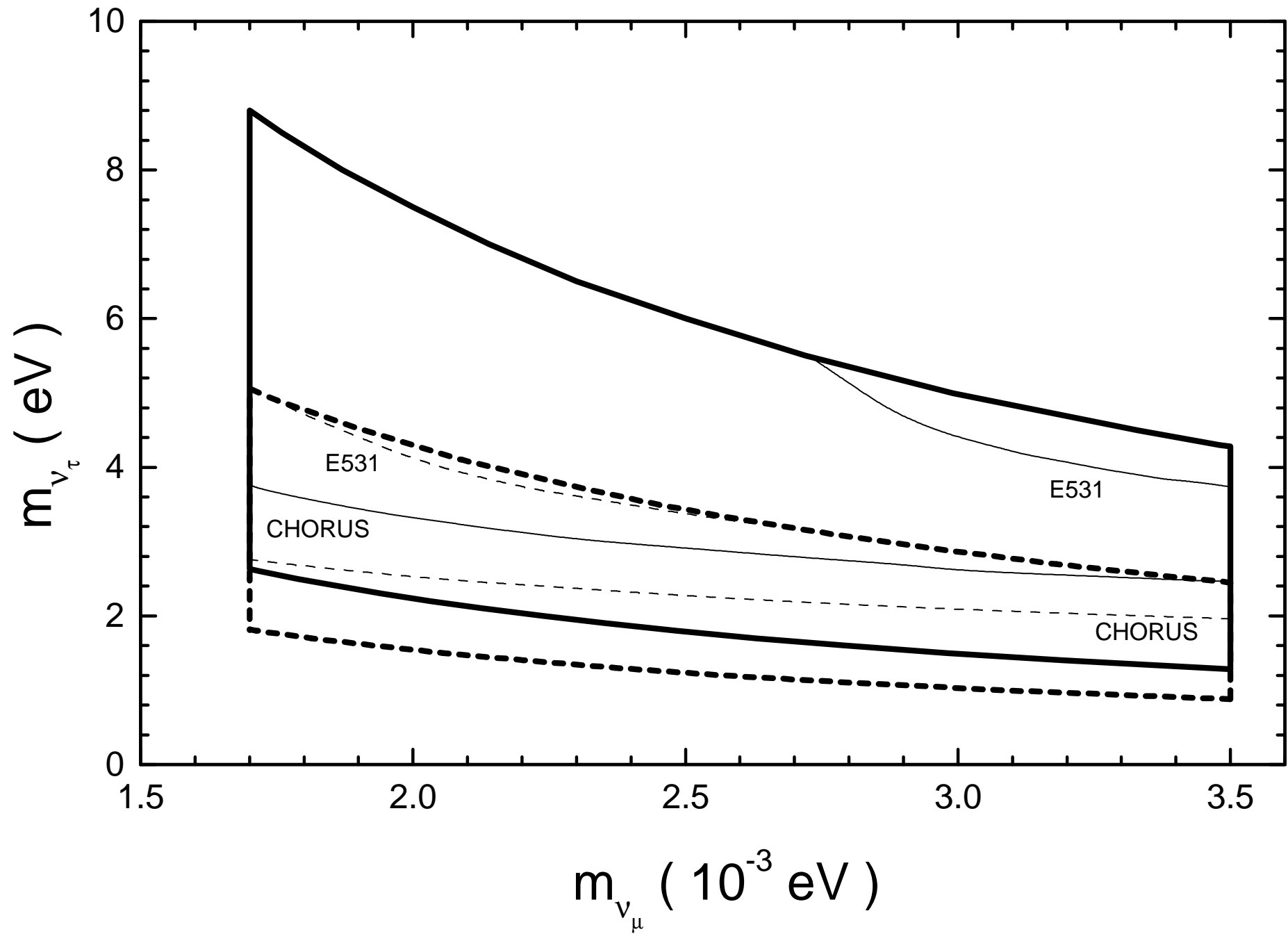
{\bf {Fig.1}.}The allowed regions in the $m_{\nu _{\mu }}$,$m_{\nu _{\tau}}$
plane.\newline

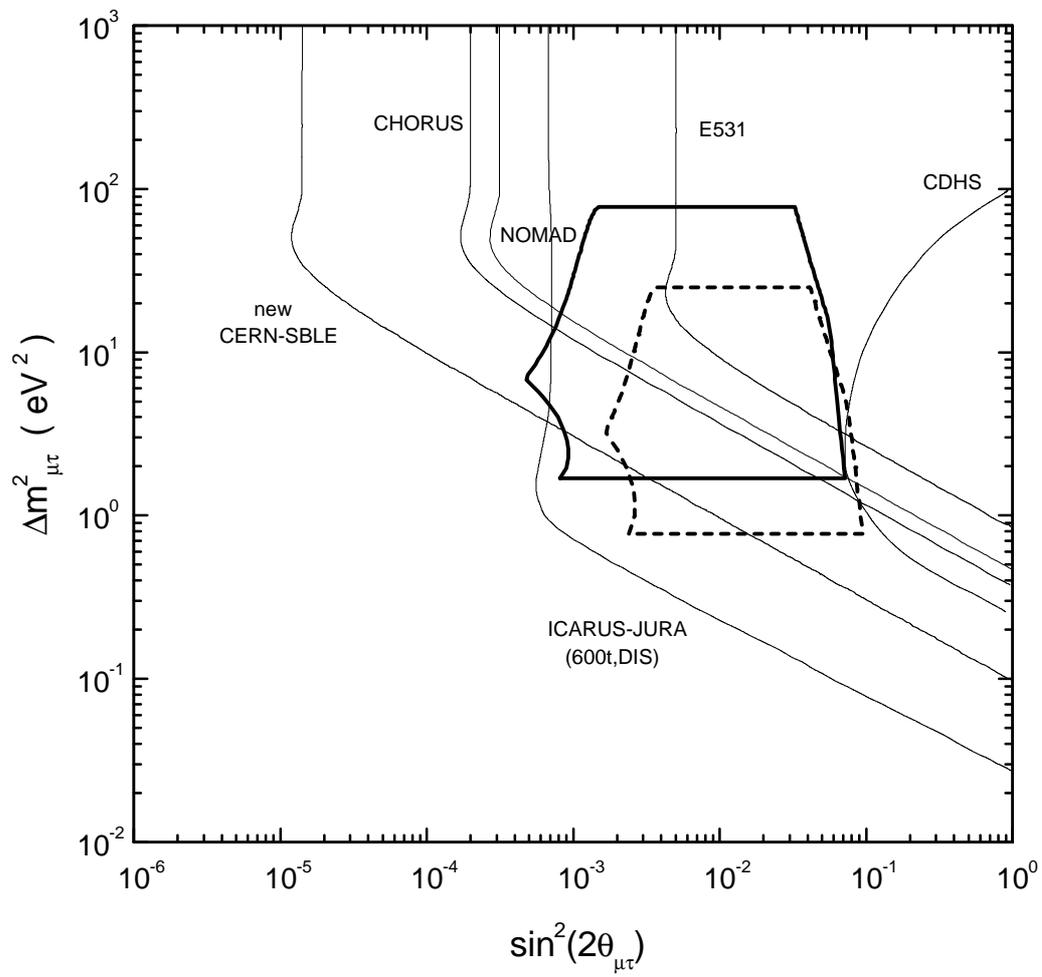
{\bf {Fig.2}.}The allowed regions in the $\nu _{\mu }$-$\nu _{\tau }$
oscillation plot.\newline

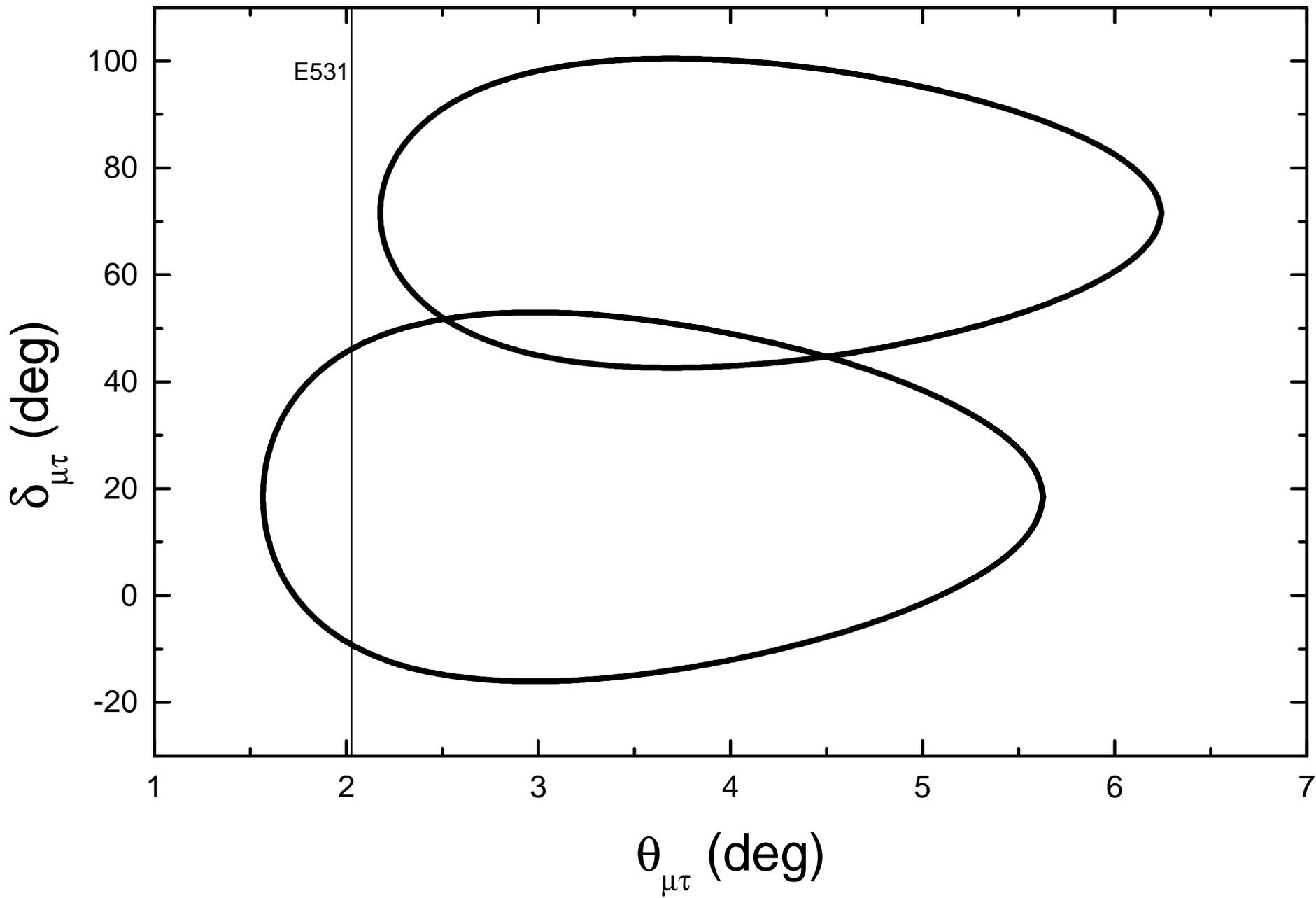
{\bf {Fig.3}.} The function $\delta _{\mu \tau }(\theta _{\mu \tau })$ for $%
M_{3}=m_{infl}/2$ and `central' values of $m_{2}$,$m_{3}$,$n_{L}/s$.


\begin{references}
\bibitem{}  For the consistency of the CHDM model, the Hubble constant is
about $50\,\,kms^{-1}Mpc^{-1}$.
\end{references}
\end{document}